\documentclass[12pt,a4paper]{article}

\usepackage{amsmath,amssymb}
\usepackage[dvips]{hyperref}

\newcommand{\di}{\genfrac{}{}{0pt}{}}
\def\under#1{\kern.4pt\underline{\kern-.4pt{}#1\kern-.4pt}\kern.4pt}
\allowdisplaybreaks[4]

\newtheorem{Theorem}{Theorem}
\newtheorem{Proposition}[Theorem]{Proposition}
\newtheorem{Lemma}[Theorem]{Lemma}
\newtheorem{Definition}[Theorem]{Definition}
\newtheorem{Corollary}[Theorem]{Corollary}

\renewcommand{\title}[1]{\vspace{10mm}\noindent{\Large{\bf #1}}\vspace{8mm}}
\newcommand{\authors}[1]{\noindent{\large #1}\vspace{3mm}}
\newcommand{\address}[1]{{\itshape #1\vspace{2mm}}}

\makeatletter
\def\section{\@startsection{section}{1}{\z@}{-3.25ex plus -1ex minus
    -.2ex}{1.5ex plus .2ex}{\normalfont\large\bfseries}}
\def\subsection{\@startsection{subsection}{1}{\z@}{-3.25ex plus -1ex
    minus -.2ex}{1.5ex plus .2ex}{\normalfont\itshape}}

\renewenvironment{thebibliography}[1]
         {\section*{References}\addcontentsline{toc}{section}{References}%
          \frenchspacing\small
          \begin{list}{[\arabic{enumi}]}
         {\usecounter{enumi}\parsep=2pt\@plus\p@\topsep 0pt
         \settowidth{\labelwidth}{[#1]}
         \leftmargin=\labelwidth\advance\leftmargin\labelsep
         \rightmargin=0pt\itemsep=0pt\sloppy}}{\end{list}}

\makeatother

\begin{document}

\begin{center}

  \title{\bf Solvable limits of a 4D noncommutative QFT}

\authors{Harald {\sc Grosse}$^1$ and Raimar {\sc Wulkenhaar}$^2$}

\address{$^{1}$\,Fakult\"at f\"ur Physik, Universit\"at Wien\\
Boltzmanngasse 5, A-1090 Wien, Austria}

\address{$^{2}$\,Mathematisches Institut der Westf\"alischen
  Wilhelms-Universit\"at\\
Einsteinstra\ss{}e 62, D-48149 M\"unster, Germany}

\footnotetext[1]{harald.grosse@univie.ac.at}
\footnotetext[2]{raimar@math.uni-muenster.de}

\vskip 1cm

\textbf{Abstract} \vskip 3mm
\begin{minipage}{13cm}%

  In previous work we have shown that the ($\theta{\to} \infty$)-limit
  of $\phi^4_4$-quantum field theory on noncommutative Moyal space is
  an exactly solvable matrix model. In this paper we translate these
  results to position space. We show that the Schwinger functions are
  symmetric and invariant under the full Euclidean group. The
  Schwinger functions only depend on matrix correlation functions at
  coinciding indices per topological sector, and clustering is
  violated.  We prove that Osterwalder-Schrader reflection positivity
  of the Schwinger two-point function is equivalent to the question
  whether the diagonal matrix two-point function is a Stieltjes
  function.  Numerical investigations suggest that this can at best be
  expected for the wrong sign of the coupling constant. The
  corresponding Wightman functions would describe particles which
  interact without momentum transfer. The theory differs from a free
  theory by the presence of non-trivial topological sectors.
\end{minipage}
\end{center}

\noindent
{\bf MSC2010:} 81T08, 81T16, 81T27, 81T75
\\[1ex]
{\bf Keywords:} quantum field theory in 4 dimensions, exactly solvable
models, Schwinger functions, Osterwalder-Schrader correspondence

\section{Summary of previous work}

Years ago we have introduced in \cite{Grosse:2004yu} a quantum field
theory model on four-dimensional Moyal space which is defined by the following 
action functional for a real scalar field $\phi$:
\begin{align}
S[\phi]=64\pi^2
\int d^4x\Big(
\frac{Z}{2} \phi (-\Delta+ \Omega_{bare}^2 \|2\Theta^{-1}x\|^2 +\mu_{bare}^2) \phi
+ \frac{\lambda_{bare} Z^2}{4} \phi\star \phi\star \phi \star \phi \Big)(x)
\;.
\label{GW}
\end{align}
Here $\Theta$ is a skew-symmetric $4\times 4$-matrix which 
defines the Moyal product 
\begin{align}
(f\star g)(x)=\int_{\mathbb{R}^d\times \mathbb{R}^d}
\frac{dy\,dk}{(2\pi)^d} 
f\big(x+\tfrac{1}{2} \Theta k\big) \,g\big(x+y\big)\, e^{\mathrm{i}
  \langle k,y\rangle}\;,\qquad f,g\in \mathcal{S}(\mathbb{R}^d)\;.
\label{Moyal}
\end{align}
We have proved in \cite{Grosse:2004yu} that the Euclidean quantum
field theory arising from (\ref{GW}) is perturbatively renormalisable.
This means the following: One introduces a momentum cut-off $\Lambda$
and normalises four (relevant and marginal) correlation functions to
$(\lambda,\Omega,1,\mu^2)$ independently of $\Lambda$. Then one proves
that the parameters $(\lambda_{bare},\Omega_{bare},Z,\mu_{bare})$ in
(\ref{GW}) are functions of $(\lambda,\Omega,\mu,\Lambda)$ in such a
way that all correlation functions of the model, considered as
functions of $(\lambda,\Omega,\mu, \Lambda)$ and as formal power
series in $\lambda$, are finite for $\Lambda\to \infty$ order by order
in $\lambda$.

A key observation is that $\Omega_{bare}=\Omega=1$ is a fixed point of
(\ref{GW}). At this fixed point a miracle occurs: As shown by
Disertori-Gurau-Magnen-Rivasseau \cite{Disertori:2006nq}
order by order in perturbation theory, $\lim_{\Lambda\to
  \infty}\lambda_{bare}(\lambda,\mu,\Lambda)$ differs from $\lambda$
only by a finite ratio, i.e.\ the $\beta$-function vanishes.  This is
in sharp contrast with usual $\phi^4_4$-model in which
$\lambda_{bare}(\lambda,\mu,\Lambda)$ develops a singularity, the
Landau pole, already at finite $\Lambda$.

Vanishing of the $\beta$-function is often a sign of integrability.
After initial steps \cite{Grosse:2009pa} which pointed into the right
direction, we have rigorously proved in \cite{Grosse:2012uv} that a
natural scaling limit of the Euclidean field theory associated with
(\ref{GW}) is exactly solvable\footnote{In fact we prove in
  \cite{Grosse:2012uv} that general quartic matrix models with action
  $S=\mathrm{tr}(E\phi^2+\frac{\lambda}{4} \phi^4)$ are exactly
  solvable.}.  Key ingredients in this proof are the formulation as a
matrix model, the use of Ward identities in Schwinger-Dyson equations
and the theory of singular integral equations of Carleman type. We
have proved that all correlation functions of the model are computable
in terms of the solution (which exists by the Schauder fixed point
theorem) of a non-linear integral equation for a smooth, positive,
monotonously decreasing function on $\mathbb{R}_+$ that vanishes with
all derivatives at $\infty$.

The passage from (\ref{GW}) to a matrix model is achieved by the expansion 
\begin{align}
\phi(x)=:\sum_{\under{m},\under{n}\in \mathbb{N}^2} \phi_{\under{m}\under{n}} 
f_{\under{m}\under{n}}(x)\;,\qquad
f_{\under{m}\under{n}}(x):=
f_{m_1n_1}(x^0,x^1) f_{m_2n_2}(x^2,x^3)\;,
\label{phi-x-fmn}
\end{align}
where $x=(x^0,x^1,x^2,x^3)\in \mathbb{R}^4$, $\under{m}=(m_1,m_2)\in
\mathbb{N}^2$ and 
\begin{align}
f_{mn}(y^0,y^1)= 
2 (-1)^m \sqrt{\frac{m!}{n!}}\Big(\sqrt{\frac{2}{\theta}} 
\|y\| e^{\mathrm{i}\eta} \Big)^{n-m} 
L^{n-m}_m\Big(\frac{2\|y\|^2)}{\theta}\Big) e^{-\frac{\|y\|^2}{\theta}}\;,
\label{fmn}
\end{align}
for $y=(y^0,y^1)\equiv \|y\| e^{i\eta} \in \mathbb{R}^2\equiv
\mathbb{C}$. The $L^\alpha_m(t)$ are associated Laguerre polynomials
of degree $m$ in $t$. After an appropriate coordinate transformation
in $\mathbb{R}^4$, the only non-vanishing components of $\Theta$ in
(\ref{Moyal}) are
$\Theta_{12}=-\Theta_{21}=\Theta_{34}=-\Theta_{43}=:\theta>0$. In this
situation the matrix basis $f_{\under{m}\under{n}}$ satisfies
$(f_{\under{k}\under{l}}\star f_{\under{m}\under{n}})(x)=\delta_{\under{m}\under{l}}
f_{\under{k}\under{n}}(x)$ and $\int_{\mathbb{R}^4}dx
\;f_{\under{m}\under{n}}(x)=(2\pi\theta)^2
\delta_{\under{m}\under{n}}$. With these identities and with
properties of the Laguerre polynomials, (\ref{GW}) takes at
$\Omega=1$, $\lambda_{bare}\equiv\lambda$ and with
$|\under{m}|:=m_1+m_2$ the form
\begin{align}
S[\phi]&=  V\Big( \sum_{\under{m},\under{n} \in \mathbb{N}^2}
E_{\under{m}} \phi_{\under{m}\under{n}} \phi_{\under{n}\under{m}} 
+ \frac{S_{int}[\phi]}{V}\Big)\;,\qquad V:=\Big(\frac{\theta}{4}\Big)^2\;,
\nonumber
\\
E_{\under{m}}
&= Z\Big( \frac{|\under{m}|}{\sqrt{V}}
+\frac{\mu_{bare}^2}{2} \Big) \;, 
\qquad
\frac{S_{int}[\phi]}{V}= \frac{Z^2\lambda}{4}
\sum_{\under{m},\under{n},\under{k},\under{l} 
\in \mathbb{N}^2 } 
\phi_{\under{m}\under{n}} \phi_{\under{n}\under{k}} 
\phi_{\under{k}\under{l}} \phi_{\under{l}\under{m}}\;.
\label{Vphi}
\end{align}

In \cite{Grosse:2012uv} we have studied the matrix representation of
the renormalised free energy density\footnote{$\mathcal{F}$ is related
  to $\mathcal{W}$ in \cite{Grosse:2012uv} by
  $\mathcal{F}[J]=\frac{1}{64\pi^2 V\mu^8}\mathcal{W}[J]$.}
\begin{align}
\mathcal{F}[J]:=\frac{1}{64\pi^2 V^2\mu^8}\log\left(
\dfrac{\int \mathcal{D}[\phi] \;
e^{-S[\phi]+ V \sum_{\under{a},\under{b}\in \mathbb{N}^2} \phi_{\under{a}\under{b}} 
J_{\under{b}\under{a}}} }{
\int \mathcal{D}[\phi] \; e^{- S[\phi]}}
\right)_{\di{Z\mu_{bare}^2\mapsto \mu^2}{Z\mapsto (1+\mathcal{Y})}}
\label{F-density}
\end{align}
in a natural scaling limit to continuous matrix indices.  Here $\mu^2$
is the renormalised squared mass. The unusual wavefunction
renormalisation $Z\mapsto (1+\mathcal{Y})$ for continuous matrix
indices simplified the resulting equations enormously, and we keep
this convention in the present paper.  By (\ref{Vphi}), the expansion
coefficients of $\mathcal{F}$ only depend on the 1-norms $|\under{m}|$
of the matrix indices so that index summations over $\under{m}$
restrict to summations over $|\under{m}|$ with measure
$|\under{m}|+1$. In \cite{Grosse:2012uv} we have introduced a cut-off
$\mathcal{N}$ in all these summations and coupled it to the volume by
\begin{equation}
\frac{\mathcal{N}}{\sqrt{V}}  = \mu^2(1+\mathcal{Y}) \Lambda^2\;,
\label{limit}
\end{equation}
where $\Lambda^2\in \mathbb{R}_+$ is an integral cut-off which at the end
is sent to $\infty$. Therefore, the coupled limit $(\mathcal{N},V)\to
\infty$ converges to a Riemann integral
\begin{align}
\lim_{(\mathcal{N},V)\to \infty} 
\frac{1}{V} \sum_{\|\under{m}\|=0}^{\mathcal{N}} (|\under{m}|+1) 
f\Big(\frac{|\under{m}|}{\sqrt{V}}\Big)
= \mu^4(1+\mathcal{Y})^2
\int_0^{\Lambda^2}
a\, da \;f\big(\mu^2(1+\mathcal{Y})a\big)\;.
\end{align}

Correlation functions in matrix models fall into topological sectors
which are distinguished by the genus $g$ of a Riemann surface and the
number $B$ of boundary components (punctures, marked points, faces) of
the surface. It turns out \cite{Grosse:2012uv} that in the scaling
limit $(\mathcal{N},V)\to \infty$ subject to (\ref{limit}), all higher
genus contributions with $g\geq 1$ are scaled away, whereas there are
reasons to keep the boundary components $B\geq 2$. Every boundary
component carries a cycle $J_{\under{M}^j}:=J_{\under{m}_1\under{m}_2}
J_{\under{m}_2\under{m}_3}\cdots J_{\under{m}_{j-1}\under{m}_j}
J_{\under{m}_j\under{m}_1}$ of external sources, where
$\under{M}^j=\under{m}_1 \dots \under{m}_j$ stands for a collection of
$j$ indices in $\mathbb{N}^2$. In these notations, the free energy
density has a decomposition
\begin{align}
\mathcal{F}[J]=\frac{1}{64\pi^2}
\sum_{K=1}^\infty \sum_{\genfrac{}{}{0pt}{1}{n_1,\dots,n_K = 0}{n_K\geq 1}}^\infty
\!\! \frac{1}{(V\mu^4)^B} 
\Big(\!\prod_{j=1}^K \frac{1}{n_j!j^{n_j}}\!\Big)
\!\!\! \sum_{\under{M}^j_{i_j} \in (\mathbb{N}^2)^j} \!\!\!\!\!
\tilde{G}_{|\under{M}_1^{1}|\dots| \under{M}_{n_1}^{1}|
\dots |\under{M}_1^{K}|\dots| \under{M}_{n_K}^{K}|}
\prod_{j=1}^K \prod_{i_j=1}^{n_j} \!\frac{J_{\under{M}_{i_j}^{j}}}{\mu^{4j}}\,.
\label{WGJ}
\raisetag{2ex}
\end{align}
The total number of $J$-cycles in a function
$\tilde{G}_{|\under{M}_1^{1}|\dots| \under{M}_{n_1}^{1}| \dots
  |\under{M}_1^{K}|\dots| \under{M}_{n_K}^{K}|}$ is its number
$B=n_1+\dots+n_K$ of boundary components.
Defining $N:=n_1+2n_2+\dots+Kn_K$ we let
$\tilde{G}_{|\under{M}_1^{1}|\dots| \under{M}_{n_1}^{1}| \dots
  |\under{M}_1^{K}|\dots| \under{M}_{n_K}^{K}|}
=: \mu^N 
G_{|\under{M}_1^{1}|\dots| \under{M}_{n_1}^{1}| \dots
  |\under{M}_1^{K}|\dots| \under{M}_{n_K}^{K}|}$, where $G$ is 
a dimensionless function as in \cite{Grosse:2012uv}.

We have shown in \cite{Grosse:2012uv} that in the scaling limit
$(V,\mathcal{N})\to \infty$ subject to (\ref{limit}), followed by the
continuum limit $\Lambda\to \infty$, the functions $\tilde{G}$ of the
matrix indices converge to functions of continuous
variables\footnote{We change the notation of \cite{Grosse:2012uv}: The
  functions of continuous variables $G_{a_1\dots | \dots | \dots a_N}$
  in \cite{Grosse:2012uv} are written as $G(a_1,\dots | \dots | \dots
  ,a_N)$ in the present paper, whereas $G_{|m_1\dots |\dots |\dots
    m_N|}$ is reserved
  for discrete matrix indices. Similarly for $\tilde{G}$.}
\begin{align}
\lim_{\Lambda\to \infty} \lim_{(V,\mathcal{N})\to \infty} 
\tilde{G}_{|\under{M}_1^{1}|\dots| \under{M}_{n_1}^{1}|
\dots |\under{M}_1^{K}|\dots| \under{M}_{n_K}^{K}|}
=: \tilde{G}\big(A_1^{1}|\dots| A_{n_1}^{1}|
\dots |A_1^{K}|\dots| A_{n_K}^{K}\big)\;,
\end{align}
with $A^j:=a_1a_2\dots a_j\in \mathbb{R}_+^j$ a cycle of continuous variables 
$a_i\in \mathbb{R}_+$ related to the discrete cycle 
$\under{M}^j\in (\mathbb{N}^2)^n$ by
\begin{align}
a_i\mu^2(1+\mathcal{Y}):= 
\frac{|\under{m}_i|}{\sqrt{V} }\;.
\label{cont-var}
\end{align}
That is, write first $\tilde{G}_{|\under{m}_1\dots|\dots|\dots
  \under{m}_N|}$ as function of $\frac{|\under{m}_i|}{\sqrt{V}}$
according to (\ref{Vphi}) and replace for the limit
$\frac{|\under{m}_i|}{\sqrt{V}} \mapsto a_i\mu^2(1+\mathcal{Y})$. The
continuous functions $\tilde{G}(A)$ then have a finite non-zero limit
$\lim_{\Lambda\to \infty} \lim_{(V,\mathcal{N})\to \infty}$ when
expressed in terms of $A$. Key result of \cite{Grosse:2012uv} was that
all these functions $\tilde{G}(a_1,\dots|\dots|\dots ,a_N)$ can be
computed explicitly in terms of the solution of the non-linear
integral equation for the boundary 2-point function
\begin{align}
G(a,0)
= \frac{1}{1+a}
\exp\Bigg(- \lambda
\int_0^a \!\! dt \int_0^\infty 
\frac{dp}{\big(\lambda\pi p \big)^2
+\big( t+ \frac{1 + \lambda\pi p
\mathcal{H}_p[G(\bullet, 0)]}{G(p,0)}\big)^2}
\Bigg) .
\label{Ga0}
\end{align}
Here, ${\mathcal{H}_a}$ is the Hilbert transform
$\displaystyle {\mathcal{H}_a}[f(\bullet)]=\frac{1}{\pi}
\lim_{\Lambda \to \infty}\lim_{\epsilon\to 0}
\Big(\int_0^{a-\epsilon}+\int_{a+\epsilon}^{\Lambda^2}\Big) dp
\;\frac{f(p)}{p-a}$. The general 2-point function is then obtained from 
\begin{align}
G(a,b)
&=
\frac{e^{\mathcal{H}_a[\vartheta_b(\bullet)]-\mathcal{H}_0[\vartheta_0(\bullet)]}}{
\sqrt{ (\lambda\pi a)^2 
+ 
\big( b{+} \frac{1 + \lambda\pi a
\mathcal{H}_a[G(\bullet, 0)]}{G(a,0)}\big)^2}}
\;,& 
\vartheta_b(p) &:=\di{\raisebox{-1.2ex}{\mbox{\normalsize$\arctan$}}}{
\mbox{\scriptsize$[0,\pi]$}}
\Bigg(\frac{ \lambda\pi p }{
b{+} \frac{1 +\lambda \pi p \mathcal{H}_p
[G(\bullet, 0)]}{G(p,0)} }\Bigg)\;.
\label{Gab}
\end{align}

The interpretation of the prefactor $\frac{1}{V^B}$ in (\ref{WGJ})
remained somewhat obscure in \cite{Grosse:2012uv}. We argued that at
this point the limit $V\to \infty$ (which would remove everything, or
rather would restrict $\mathcal{W}=V\mathcal{F}$ to the sector $B=1$)
should not be taken.  In this paper we confirm this by showing that
connected Schwinger functions in position space contribute a factor
$V$ per boundary component.

\section{Schwinger functions}

\begin{Definition}
\label{Def:Schwinger}
The connected $N$-point Schwinger function associated
with the action (\ref{GW}) is \underline{defined} as
\begin{align}
\mu^N S_c(\mu x_1,\dots ,\mu x_N) := \lim_{V\to \infty} \!\!\!\!
\sum_{\under{m}_1,\under{n}_1,\dots,
\under{m}_N,\under{n}_N \in \mathbb{N}^2} 
\!\!\!\!\!\!\!\! f_{\under{m}_1\under{n}_1}(x_1) 
\cdots f_{\under{m}_N\under{n}_N}(x_N) 
\frac{\mu^{4N} \partial^N \mathcal{F}[J]}{\partial J_{\under{m}_1\under{n}_1}
\dots \partial J_{\under{m}_N\under{n}_N}}\bigg|_{J=0}\;.
\label{Schwinger}
\end{align}
\end{Definition}
This definition requires some explanation:
\begin{itemize} \itemsep 0pt
\item Schwinger functions are often introduced as the connected part of 
\begin{align}
\langle \varphi(x_1)\cdots \varphi(x_N)\rangle 
= 
\dfrac{\int \mathcal{D}[\phi] \;
\phi(x_1)\cdots \phi(x_N) e^{-S[\phi]} }{
\int \mathcal{D}[\phi] \; e^{- S[\phi]}}\;.
\label{PI}
\end{align}
The connected part can be expressed as functional derivative of $\log
\mathcal{Z}[J]$ with respect to sources $J(x_1),\dots, J(x_N)$, for
$\mathcal{Z}[J]:= \int \mathcal{D}[\phi] \; e^{- S[\phi]+\int dx\;
\phi(x)J(x)}$.  However, this can only make sense in a renormalisation
prescription, and renormalisation involves discussion of the infinite
volume limit. One typically finds that $\log \mathcal{Z}[J]$ is proportional
to the volume so that (\ref{PI}) does not make sense for $V\to
\infty$. Accordingly, (\ref{Schwinger}) does not agree with the
connected part of $\langle \varphi(x_1)\cdots \varphi(x_N)\rangle$.

Our point of view is that only the free energy density $\mathcal{F}$
makes sense in the infinite volume limit, and that \emph{Schwinger
  functions are densities}, too.  This implies that all quantities of
the renormalised theory must be viewed as dimensionless ratios with
an appropriate power of the mass scale $\mu$. In particular, $V\to
\infty$ means $(V\mu^4)\to \infty$ with $\mu$ fixed. Absolute
positions $x$ loose their meaning in the limit $V\to \infty$; only
$\mu x$ is meaningful.

\item The definition (\ref{Schwinger}) involves a non-na\"{\i}ve
  wavefunction renormalisation. From (\ref{F-density}) we would expect
  that a Schwinger function which represents $\langle
  \varphi(x_1)\cdots \varphi(x_N)\rangle_c$ is the derivative
  $\big(\frac{1}{V}\frac{\partial}{\partial J}\big)^N$ applied to
  $\mathcal{F}[J]$, and not $\big(\mu^4\frac{\partial}{\partial
    J}\big)^N$ as imposed in (\ref{Schwinger}). It is not difficult to
  see that the removal of the factor $(V\mu^4)^{-1}$ corresponds
  precisely to a wavefunction renormalisation $\sqrt{Z}\mapsto
  \frac{\sqrt{Z}}{V\mu^4}$.

\item 
The mass scale is introduced by the normalisation
  $\tilde{G}_{\under{0}\under{0}}=\mu^2$. For the free theory
  $\lambda=0$ in (\ref{Vphi}) we can compute the free energy density exactly:
\begin{align*}
\mathcal{F}[J]\big|_{\lambda=0}
&=\frac{1}{64\pi^2} \cdot \frac{1}{V\mu^4}
\cdot \frac{1}{2} \cdot \sum_{\under{m},\under{n}\in \mathbb{N}^2}
\Big(\frac{\mu^4}{Z(\frac{|m|+|n|}{\sqrt{V}}+\mu_{bare}^2)}\Big) 
\frac{J_{\under{m}\under{n}}}{\mu^4}\frac{J_{\under{n}\under{m}}}{\mu^4}
\\
&=\frac{1}{64\pi^2} \cdot \frac{1}{V\mu^4}
\cdot \frac{1}{2} \cdot \sum_{\under{m},\under{n}\in \mathbb{N}^2}
\Big(\frac{\mu^4}{(V\mu^4)^2Z(\frac{|m|+|n|}{\sqrt{V}}+\mu_{bare}^2)}\Big) 
(VJ_{\under{m}\under{n}})(VJ_{\under{n}\under{m}})\;.
\end{align*}
This shows again that (\ref{Schwinger}) is compatible with $Z=1$ and
$\mu_{bare}^2=\mu^2$ for the free theory, and that the na\"{\i}vely
expected $\big(\frac{1}{V}\frac{\partial}{\partial J}\big)^N$ leads to
the same result if we let $\sqrt{Z}=\frac{1}{V\mu^4}$ and
$\mu_{bare}^2=\mu^2$. 

\item Since the Gau\ss{}ian in (\ref{fmn}) distinguishes the origin
  and the decomposition $f_{\under{m}\under{n}}(x)=
  f_{m_1n_1}(x^0,x^1) f_{m_2n_2}(x^2,x^3)$ distinguishes pairs of
  coordinate directions, the Schwinger functions (\ref{Schwinger})
  are, \emph{a priori}, only invariant under the subgroup $SO(2)\times SO(2)$
  of the Euclidean group $\mathbb{R}^4 \rtimes SO(4)$. 

\item In contrast, the Schwinger functions are \emph{fully symmetric 
 in all its arguments}.

\end{itemize}
The differentiation of (\ref{WGJ}) with respect to the $J$'s in
(\ref{Schwinger}) is a standard combinatorial problem. For 
$\under{M}^j=\under{m}_1 \dots \under{m}_j$ we define 
\begin{align}
f_{\under{M}^j}(x_1,\dots,x_j) :=
f_{\under{m}_1\under{m}_2}(x_1)
f_{\under{m}_2\under{m}_3}(x_2)\cdots f_{\under{m}_{j-1}\under{m}_j}(x_{j-1})
f_{\under{m}_j\under{m}_1}(x_j)\;.
\end{align}
In terms of dimensionless functions
$G_{|\under{M}_1^{1}|\dots| \under{M}_{n_1}^{1}|
\dots |\under{M}_1^{K}|\dots| \under{M}_{n_K}^{K}|}
= \mu^{-N} \tilde{G}_{|\under{M}_1^{1}|\dots| \under{M}_{n_1}^{1}|
\dots |\under{M}_1^{K}|\dots| \under{M}_{n_K}^{K}|}$ for
$N:=n_1+2n_2+\dots+ K n_K$  we have
\begin{align}
&S_c(\mu x_1,\dots,\mu x_N)
\nonumber
\\*[-1ex]
&= \lim_{V\to \infty} \frac{1}{64\pi^2}
\sum_{n_1+2n_2+\dots+ K n_K=N}
\frac{1}{(V\mu^4)^{n_1+n_2+\dots+n_K}} 
\sum_{\under{M}^j_{i_j} \in (\mathbb{N}^2)^j}
G_{|\under{M}_1^{1}|\dots| \under{M}_{n_1}^{1}|
\dots |\under{M}_1^{K}|\dots| \under{M}_{n_K}^{K}|}
\nonumber
\\*[-2ex]
& \times 
\sum_{\sigma \in \mathcal{S}_N} \Big(\prod_{j=1}^K \frac{1}{j^{n_j}}\Big)
\big(f_{\under{M}_1^1}(x_{\sigma(1)})
\cdots f_{\under{M}_{n_1}^1}(x_{\sigma(n_1)})\big)
\nonumber
\\*
& \qquad\times \big(f_{\under{M}_1^2}(x_{\sigma(n_1+1)},x_{\sigma(n_1+2)})
\cdots f_{\under{M}_{n_2}^2}(x_{\sigma(n_1+2n_2-1)},x_{\sigma(n_1+2n_2)})
\big)
\nonumber
\\*
&\qquad \times \cdots \times
\big(f_{\under{M}_1^K}(x_{\sigma(n_1+\dots+(K-1)n_{K-1}+1)},\dots 
x_{\sigma(n_1+\dots+(K-1)n_{K-1}+K)})
\cdots 
\nonumber
\\*
&\hspace*{6cm} \times 
f_{\under{M}_{n_K}^K}(x_{\sigma(n_1+\dots+Kn_K-(K-1)},x_{\sigma(n_1+\dots + Kn_K)})\;.
\label{Schwinger-1}
\end{align}
The summation is over all permutations in the symmetric group
$\mathcal{S}_N$. It is much more convenient to write
$N=j_1+\dots+j_B$, where $j_\beta$ is the length of the
$\beta^{\mathrm{th}}$ cycle in $G_{|\under{M}_1^{1}|\dots|
  \under{M}_{n_1}^{1}| \dots |\under{M}_1^{K}|\dots|
  \under{M}_{n_K}^{K}|}$. The cycles $(1,2,\dots,\beta-1)$ contain
$N_\beta^-:=j_1+\dots+j_{\beta-1}$ of the $N$ indices, and the
$\beta^{\mathrm{th}}$ cycle adds $j_\beta$ more.  With these
conventions (\ref{Schwinger-1}) can be written equivalently as
\begin{align}
S_c(\mu x_1,\dots,\mu x_N)
= \lim_{V\to \infty}
\frac{1}{64\pi^2} & \sum_{j_1+\dots+j_B=N}
\sum_{\{\under{M}_\beta^{j_\beta} \in (\mathbb{N}^2)^{j_\beta}\}_{\beta=1}^B}
G_{|\under{M}_1^{j_1}|\dots| \under{M}_{B}^{j_B}|}
\nonumber
\\*[-1ex]
& \times 
\sum_{\sigma \in \mathcal{S}_N} \prod_{\beta=1}^B
\frac{1}{V\mu^4j_\beta} f_{\under{M}_{\beta}^{j_\beta}}(x_{\sigma(N_\beta^-+1)},\dots, 
x_{\sigma(N_\beta^-+j_\beta)})\;.
\label{Schwinger-2}
\end{align}

To proceed we assume that $G_{|\under{M}_1^{1}|\dots|
  \under{M}_{B}^{j_B}|}$ has, for each cycle
$\under{M}_\beta^{j_\beta}=\under{m}^{\beta}_1
\dots\under{m}^\beta_{j_\beta}$, a representation as Laplace transform
in the total norm $\|\under{M}_\beta^{j_\beta}\|:=
|\under{m}^{\beta}_1|+\dots+|\under{m}^{\beta}_{j_\beta}|$ and Fourier
transform in neighboured differences. For
$\under{\omega}^{j}=(\omega^{j}_1,\dots\omega^{j}_{j-1}) \in
\mathbb{R}^{j-1}$ let $\langle \under{\omega}^{j}, \under{M}^j\rangle
:= \sum_{i=1}^{j-1}\omega_i(|\under{m}_{i}|-|\under{m}_{i+1}|)$. We
assume
\begin{align}
G_{|\under{M}_1^{j_1}|\dots|
  \under{M}_{B}^{j_B}|}
=\int_{\mathbb{R}_+^B} \!\!\! d(t_1,\dots, t_B) 
\int_{\mathbb{R}^{N-B}} \!\!\! 
d(\under{\omega}_1^{j_1},\dots,\under{\omega}_B^{j_B})\;
&\mathcal{G}(t_1,\under{\omega}^{j_1}_1|\dots|t_B,\under{\omega}^{j_B}_B)
\nonumber
\\*[-2ex]
&\times \prod_{\beta=1}^B e^{-\frac{t_\beta}{\sqrt{V}}\|\under{M}_\beta^{j_\beta}\|
+\frac{\mathrm{i}}{\sqrt{V}}
\langle \under{\omega}_\beta^{j_\beta},\under{M}_\beta^{j_\beta}\rangle}
\;.
\label{Laplace-Fourier}
\end{align}
The existence assumption of the inverse Laplace transform 
$\mathcal{G}$ is only a
technical trick which in the end will be reverted. 
Insertion into (\ref{Schwinger-2}) gives
\begin{align}
&S_c(\mu x_1,\dots,\mu x_N)
\nonumber
\\&= \lim_{V\to \infty}\frac{1}{64\pi^2}
\sum_{j_2+\dots+ j_B=N}
\int_{\mathbb{R}_+^B} \!\!\! d(t_1,\dots, t_B) 
\int_{\mathbb{R}^{N-B}} \!\!\! 
d(\under{\omega}_1^{j_1},\dots,\under{\omega}_B^{j_B})\;
\mathcal{G}(t_1,\under{\omega}^{j_1}_1|\dots|t_B,\under{\omega}^{j_B}_B)
\nonumber
\\*
& \times 
\sum_{\sigma \in \mathcal{S}_N} 
\prod_{\beta=1}^B \bigg(\frac{1}{V\mu^4j_\beta} \!\!\! 
\sum_{\under{M}_{\beta}^{j_\beta} \in  (\mathbb{N}^2)^{j_\beta}}
f_{\under{M}_{\beta}^{j_\beta}}(x_{\sigma(N_\beta^-+1)},\dots, 
x_{\sigma(N_\beta^-+j_\beta)})\;
e^{-\frac{t_\beta}{\sqrt{V}}\|\under{M}_\beta^{j_\beta}\|
+\frac{\mathrm{i}}{\sqrt{V}}
\langle \under{\omega}_\beta^{j_\beta},\under{M}_\beta^{j_\beta}\rangle}
\bigg)
\;.
\raisetag{2ex}
\label{Schwinger-3}
\end{align}
The index sum is achieved by Corollary~\ref{Cor:Laguerre} which we
prove in the Appendix. According to (\ref{phi-x-fmn}) the
$f_{\under{m}\under{n}}$ that we need in (\ref{Schwinger-3}) are
products of each two $f_{m_im_{i+1}}$, one for $(x_i^0,x_i^1)$, the
other for $(x_i^0,x_i^1)$. Both have in the notation of
(\ref{sum-Laguerre}) common factors
\[
z_1=e^{-\frac{t_\beta}{\sqrt{V}}+\frac{\mathrm{i}}{\sqrt{V}}
\omega_{\beta,1}},
\quad 
z_i=e^{-\frac{t_\beta}{\sqrt{V}}+\frac{\mathrm{i}}{\sqrt{V}}
(\omega_{\beta,i}-\omega_{\beta,i-1})}~\text{for }2\leq i\leq j_\beta-1\;,
\quad 
z_{j_\beta}=e^{-\frac{t_\beta}{\sqrt{V}}-\frac{\mathrm{i}}{\sqrt{V}}
\omega_{\beta,j_\beta-1}}.
\]
For $V\to \infty$ all $z_i$ converge to $1$. The denominator
$1-\prod_{i=1}^{j_\beta}(-z_i)=1-(-1)^{j_\beta}
e^{-j_\beta\frac{t_\beta}{\sqrt{V}}}$ in (\ref{sum-Laguerre})
converges to $2$ for $j_\beta$ odd but behaves as
$j_\beta\frac{t}{\sqrt{V}}$ for $j_\beta$ even.  The scalar and vector
products in (\ref{sum-Laguerre}) 
receive common factors for $V\to \infty$, so that we conclude
with $\theta=4\sqrt{V}$:
\begin{align}
&\lim_{V\to \infty} 
\frac{1}{V\mu^4j_\beta} \!\!\! 
\sum_{\under{M}_{\beta}^{j_\beta} \in  (\mathbb{N}^2)^{j_\beta}}
f_{\under{M}_{\beta}^{j_\beta}}(x_{\sigma(N_\beta^-+1)},\dots, 
x_{\sigma(N_\beta^-+j_\beta)})\;
e^{-\frac{t_\beta}{\sqrt{V}}\|\under{M}_\beta^{j_\beta}\|
+\frac{\mathrm{i}}{\sqrt{V}}
\langle \under{\omega}_\beta^{j_\beta},\under{M}_\beta^{j_\beta}\rangle}
\nonumber
\\*[-1.5ex]
&=\left\{
\begin{array}{cl}
\displaystyle 
\frac{4^{j_\beta}}{\mu^4j_\beta^3 t^2} 
\exp\bigg(-\frac{\big\|\sum_{i=1}^{j_\beta} (-1)^{i-1} x_{\sigma(N_\beta^-+i)}
\big\|^2}{2j_\beta t}\bigg) &\qquad \text{for  $j_\beta$ even,}
\\[2ex] 0 &\qquad   \text{for  $j_\beta$ odd.}
\end{array}\right.
\end{align}
The surviving factor for $j_\beta$ even can be written as
\begin{align}
&\frac{4^{j_\beta}}{\mu^4j_\beta^3 t^2} \:
e^{\bigg(-\frac{\big\|\sum_{i=1}^{j_\beta} (-1)^{i-1} x_{\sigma(N_\beta^-+i)}
\big\|^2}{2j_\beta t}\bigg)}
= \frac{4^{j_\beta}}{4\pi^2 j_\beta}
\int_{\mathbb{R}^4} \!\!\! \frac{dp_\beta}{\mu^4} \;e^{\mathrm{i}\big\langle 
p_\beta,\sum_{i=1}^{j_\beta} (-1)^{i-1} x_{\sigma(N_\beta^-+i)}\big\rangle}\,
e^{-\frac{j_\beta}{2} \|p_\beta\|^2 t}\;.
\raisetag{2ex}
\label{pbeta}
\end{align}
This result is inserted back into (\ref{Schwinger-3}). Comparing the
resulting $(t,\under{\omega})$-integral with (\ref{Laplace-Fourier}),
the $\under{\omega}$-independence of (\ref{pbeta}) forces all of the
$j_\beta$ matrix indices $m_{\beta,i}$ in $\under{M}^{j_\beta}_\beta$
to be equal.  Then $\|\under{M}^{j_\beta}_\beta\|$ is $j_\beta$ times
the norm $\|m_{\beta,i}\|$ of any of these indices. The result is
precisely the Laplace transform of $\mathcal{G}(\dots,t_\beta,\dots)$
to $G(\dots,\frac{\|m_{\beta,i}\|}{\sqrt{V}}
\mapsto\frac{\|p_\beta\|^2}{2},\dots)$.  The remaining limit $V\to
\infty$ applies to the reconstructed $G$, but as noted in
(\ref{cont-var}), this limit sends $\frac{\|m_{\beta,i}\|}{\sqrt{V}}
=\frac{\|p_\beta\|^2}{2}$ into the continuous variable
$a_i\mu^2(1+\mathcal{Y})$, hence $a_i$ into
$\frac{\|p_\beta\|^2}{2\mu^2(1+\mathcal{Y})}$. 
We have thus proved:
\begin{Proposition}
The connected Schwinger functions (\ref{Schwinger}) take the form
\begin{align}
S_c(\mu x_1,\dots,\mu x_N)
&= \frac{1}{64\pi^2}
\sum_{\di{j_1+\dots+j_B=N}{j_\beta\,\mathrm{even}}} 
\sum_{\sigma \in \mathcal{S}_N} 
\bigg(\prod_{\beta=1}^B \frac{4^{j_\beta}}{j_\beta} 
\int_{\mathbb{R}^4} \frac{d^4p_\beta}{4\pi^2\mu^4} 
\;e^{\mathrm{i}\big\langle 
\frac{p_\beta}{\mu},\sum_{i=1}^{j_\beta} (-1)^{i-1} \mu 
x_{\sigma(N_\beta^-+i)}\big\rangle}\bigg)
\nonumber
\\*[-0.5ex]
&\qquad\quad \times 
G\Big(\underbrace{\tfrac{\|p_1\|^2}{2\mu^2(1+\mathcal{Y})},\cdots,
\tfrac{\|p_1\|^2}{2\mu^2(1+\mathcal{Y})}}_{j_1}\big| \dots \big|
\underbrace{\tfrac{\|p_B\|^2}{2\mu^2(1+\mathcal{Y})},\cdots,
\tfrac{\|p_B\|^2}{2\mu^2(1+\mathcal{Y})}}_{j_B}\Big)
\;.
\label{Schwinger-final}
\end{align}
Consequently, Schwinger functions are invariant under the full
Euclidean group.  The Schwinger functions only detect the restricted
sector of the underlying matrix model where all matrix indices of a
boundary component coincide. \hfill $\square$%
\end{Proposition}

In particular, the Schwinger two-point function reads
\begin{align}
S_c(\mu x,\mu y)=\int_{\mathbb{R}^4}\frac{d^4p}{(2\pi\mu)^4} \;
e^{\mathrm{i}\langle \frac{p}{\mu},(\mu x-\mu y)\rangle} 
G\Big(\frac{\|p\|^2}{2\mu^2(1+\mathcal{Y})},
\frac{\|p\|^2}{2\mu^2(1+\mathcal{Y})}\Big)\;.
\label{S-2}
\end{align}
The perturbative result
$G(a,b)=\frac{1}{1+(1+\mathcal{Y})(a+b)}+\mathcal{O}(\lambda)$
obtained in \cite{Grosse:2012uv} agrees with the expectation
$\displaystyle 
\mu^2 S_c(\mu x,\mu y)= \int_{\mathbb{R}^4}\frac{d^4p}{(2\pi)^4} \;
\frac{e^{\mathrm{i}\langle
    p,(x-y)\rangle}}{\|p\|^2+\mu^2}+\mathcal{O}(\lambda)$
for the Euclidean $\phi^4_4$-model.

\section{Analytic continuation to Minkowski space}

\label{sec:continuation}

Under a set of conditions established by Osterwalder-Schrader
\cite{Osterwalder:1973dx, Osterwalder:1974tc}, Schwinger functions of
a Euclidean quantum field theory have an analytic continuation to
Wightman functions \cite{Streater:1964??} of a relativistic quantum
field theory. Whether this is the case for the Schwinger functions
(\ref{Schwinger-final}) is of great interest, because non-trivial
four-dimensional examples are rare.

The relation between Euclidean and Minkowskian Moyal-deformed field
theories has already been addressed in literature. In joint work of
one of us (HG) with Lechner, Ludwig and Verch \cite{Grosse:2011es} it
was proved for degenerate deformations where time remains commutative
that the Osterwalder-Schrader correspondence commutes (up to
isomorphism) with Moyal deformation. This result was achieved in an
algebraic approach to the Osterwalder-Schrader reconstruction theorem
which is due to Schlingemann \cite{Schlingemann:1998cw}. For
deformations of full rank such a correspondence cannot be expected. As
shown by Bahns \cite{Bahns:2009iq}, Wightman functions in a
Minkowskian Moyal-deformed field theory admit an analytic continuation
to imaginary time, but this continuation does not agree with the
Schwinger functions of Euclidean Moyal-deformed field theory, at least
in a framework close to perturbation theory. Adding the harmonic
oscillator potential (\ref{GW}) to the Moyal deformation is also problematic in
Minkowski space \cite{Zahn:2010yt}.

We will show in this section that the limit $\theta\to\infty$ cures
all problems \cite{Bahns:2009iq, Zahn:2010yt} arising in full-rank
Minkowskian Moyal-deformed theories. The question whether or not the
Schwinger functions (\ref{Schwinger-final}) define a Wightman quantum
field theory is, in principle, decidable in view of their exact
solution established in \cite{Grosse:2012uv}. The lack of better
knowledge of the properties of the fixed point solution (\ref{Ga0})
forces us to postpone the answer. We will extract that a necessary
condition for (\ref{Schwinger-final}) defining a Wightman theory is
that $a\mapsto G(a,a)$ is a Stieltjes function \cite{Widder:1938??}.
First numerical investigations \cite{GW-numerical} of (\ref{Ga0})
suggest that this can only be expected for the wrong sign $\lambda
\leq 0$ of the coupling constant. A related consequence of the
numerical behaviour is the negative anomalous dimension
$\eta=-2\lambda$.  This is a surprising result which in view of
\cite{Schlingemann:1998cw} shows that the Euclidean operator algebra
generated by the Schwinger function is highly sensitive to the sign of
$\lambda$. In particular, the Osterwalder-Schrader correspondence of
Moyal $\phi^4_4$-theory is inaccessible in perturbation theory.

\smallskip

Prior to analytic continuation is the expression of the Schwinger
functions in terms of position differences $\xi_{(k)}:=x_{k+1}-x_k$
\cite{Osterwalder:1974tc}. Fixing a permutation $\sigma$ and the
number $B$ of boundary components, the Schwinger functions
(\ref{Schwinger-final}) only depend on the $B$ sums
$\xi_{\beta(\sigma)}:= \sum_{l=1}^{\frac{j_\beta}{2}}
\xi_{(\sigma(N_\beta^-+2l-1))}$. We distinguish temporal and spatial
directions,
$\xi_{\beta(\sigma)}=(\xi_{\beta(\sigma)}^0,\vec{\xi}_{\beta(\sigma)})$,
$q_\beta=(q_\beta^0,\vec{q}_\beta)$ and collect these by bold symbols
$\boldsymbol{\xi_{(\sigma)}}=(\xi_{1(\sigma)},\dots,\xi_{B(\sigma)})$,
$\boldsymbol{q}=(q_1,\dots,q_B)$,
$\boldsymbol{\xi_{(\sigma)}^0}=(\xi_{1(\sigma)}^0,\dots,\xi_{B(\sigma)}^0)$,
$\boldsymbol{q^0}=(q_1^0,\dots,q_B^0)$,
$\boldsymbol{\vec{\xi}_{(\sigma)}}=(\vec{\xi}_{1(\sigma)},\dots,
\vec{\xi}_{B(\sigma)})$,
$\boldsymbol{\vec{q}}=(\vec{q}_1,\dots,\vec{q}_B)$. We also let
$\boldsymbol{\xi_{(\sigma)}^0}\cdot \boldsymbol{q^0}=\sum_{\beta=1}^B
\xi_{\beta(\sigma)}^0 q_\beta^0$ and
$\boldsymbol{\vec{\xi}_{(\sigma)}}\cdot
\boldsymbol{\vec{q}}=\sum_{\beta=1}^B \langle
\vec{\xi}_{\beta(\sigma)}, \vec{q}_\beta\rangle$.  Under appropriate
analyticity conditions \cite{Osterwalder:1974tc}, the Schwinger
functions are Fourier-Laplace transforms
\begin{align}
S_c(\mu x_1,\dots,\mu x_N)&=\sum_{\sigma \in \mathcal{S}_N}\sum_{B=1}^{\frac{N}{2}}
S_N^{\sigma,B}(\mu\boldsymbol{\xi_{(\sigma)}})\;,\nonumber
\\*
S_N^{\sigma,B}(\mu\boldsymbol{\xi_{(\sigma)}})\Big|_{\xi_{\beta(\sigma)}^0>0}
&= \frac{1}{\mu^{4B}} 
\int_{\mathbb{R}_+^B} d^B \boldsymbol{q^0} \int_{\mathbb{R}^{3B}}
d^{3B} \boldsymbol{\vec{q}} \;\;
\hat{W}^B_N(\tfrac{\boldsymbol{q}}{\mu})\;
e^{-\boldsymbol{\xi_{(\sigma)}^0}\cdot \boldsymbol{q^0}
+\mathrm{i}\boldsymbol{\vec{\xi}_{(\sigma)}}\cdot
\boldsymbol{\vec{q}}}\;.
\label{Fourier-Laplace}
\end{align}
The functions $\hat{W}^B_N$ on $\mathbb{R}^{4B}$ are candidates for the
Fourier transform of Wightman $N$-point functions with $B$ independent
position differences. We remark that this restricted position
dependence is for $B=\frac{N}{2}$ identical with a free field theory
where the Osterwalder-Schrader reconstruction theorem is established.

Proving the Osterwalder-Schrader axioms \cite{Osterwalder:1974tc} for
the Schwinger functions (\ref{Schwinger-final}), which would imply
(\ref{Fourier-Laplace}) with the correct properties of $\hat{W}^B_N$,
is an open problem which we only address partly. We restrict ourselves
to the 2-point function (\ref{S-2}) which has the usual number $B=1$
of independent position difference vectors. We prove:

\begin{Proposition}
\label{Prop:Stieltjes}
Necessary and sufficient for the Schwinger 2-point function $S_2(\mu
\xi):=\sum_{\sigma \in \mathcal{S}_2} S_2^{\sigma,1}(\mu\xi)$ being
the Fourier-Laplace transform of a positive Wightman function
$\hat{W}^1_2(\frac{q}{\mu})$ is that $a \mapsto G(a,a)$ is a Stieltjes
function,
\begin{align}
G(a,a)= \int_0^\infty \frac{d \rho(\frac{M^2}{\mu^2})}{
(2(1+\mathcal{Y})a +\frac{M^2}{\mu^2})}\;, 
\label{G-Stieltjes}
\end{align}
where $\rho(\frac{M^2}{\mu^2})$ is a positive measure.
\end{Proposition}
Stieltjes functions were thoroughly studied by Widder 
\cite{Widder:1938??}: A function $\mathbb{R}_+\ni x\mapsto f(x) \in
\mathbb{R}$ is Stieltjes iff $f$ is smooth and
\begin{enumerate}
\item[(S1)] $f(x) \geq 0$ for all $x\in \mathbb{R}_+$
\item[(S2)] $(-1)^n \dfrac{d^{2n+1}}{dx^{2n+1}} \big(x^{n+1}f(x)\big) \geq 0
$ for all $x\in \mathbb{R}_+$ and $n\in \mathbb{N}$.
\end{enumerate}

\medskip

\noindent
\emph{Proof of Prop.~\ref{Prop:Stieltjes}.}  This is a consequence of
the K\"all\'en-Lehmann spectral representation \cite{Kallen:1952zz,
  Lehmann:1954xi}.  Inserting (\ref{G-Stieltjes}) into (\ref{S-2}) we
have 
\begin{align}
S_2(\mu \xi):= \sum_{\sigma \in \mathcal{S}_2} S^{\sigma,1}_2 (\mu\xi_\sigma) 
=\int_0^\infty d\rho (\tfrac{M^2}{\mu^2}) 
\int_{\mathbb{R}^4} \frac{dp^0 d^3 \vec{p}}{(2\pi\mu)^4}
\frac{\mu^{2} e^{\mathrm{i} p^0\xi^0+\mathrm{i} \vec{p}\cdot \vec{\xi} }}{(
  (p^0)^2 + \vec{p}\cdot \vec{p}+M^2)}\;.
\end{align}
For $\xi^0>0$ the $p^0$-integral is evaluated by the residue theorem:
\begin{align}
S_2(\mu \xi)\Big|_{\xi^0>0} {=}\int_0^\infty \!\!\!
d\rho (\tfrac{M^2}{\mu^2}) 
\int_{\mathbb{R}^3} \! \frac{d^3 \vec{p}}{(2\pi\mu )^3}
\frac{\mu}{2\omega_{\vec{p},M}}
e^{- \omega_{\vec{p},M} \xi^0+\mathrm{i} \vec{p}\cdot \vec{\xi} }\;,
\quad \omega_{\vec{p},M}:= \sqrt{\vec{p}\cdot \vec{p}+M^2}\;.
\end{align}
This gives the desired representation 
$\displaystyle S_2(\mu \xi)\Big|_{\xi^0>0} =\frac{1}{\mu^4}
\int_0^\infty \!\!\! dq^0
\int_{\mathbb{R}^3} \!\! d\vec{q} \;\;\hat{W}^1_2\big(\tfrac{q}{\mu}\big) 
\;e^{-q^0\xi^0 +\mathrm{i}  \vec{q}\cdot \vec{\xi}}$ as
Fourier-Laplace transform with
\begin{align}
\hat{W}^1_2(q)&= \frac{\theta(q^0)}{(2\pi)^3} \int_0^\infty 
d\rho(\tfrac{M^2}{\mu^2}) \;
\delta\Big(\frac{(q^0)^2-\vec{q}\cdot \vec{q}-M^2}{\mu^2}\Big) \;,
\label{W-2}
\end{align}
where $\theta$ and $\delta$ are the Heaviside and Dirac distributions.
The final formula (\ref{W-2}) is recognised as the K\"all\'en-Lehmann
spectral representation \cite{Kallen:1952zz, Lehmann:1954xi} of a
two-point function in a general Wightman quantum field theory. The
converse steps starting with (\ref{W-2}) show that the Stieltjes
property (\ref{G-Stieltjes}) is necessary.  \hfill $\square$%

\bigskip

We are currently unable to determine whether the matrix 2-point
function $a\mapsto G(a,a)$ is Stieltjes, i.e.\ satisfies Widder's
conditions (S1)+(S2). Every Stieltjes function is completely monotonic,
i.e.\  
\begin{enumerate}
\item[(CM)$\!\!\!\!$] $\quad (-1)^n f^{(n)}(x) \geq 0$ for all $x\in
  \mathbb{R}_+$.
\end{enumerate}
Complete monotonicity (CM) for the function $a\mapsto G(a,0)$ might be
in reach. We recall from \cite{Grosse:2012uv} that $G(a,0)$ is the
solution (\ref{Ga0}) of a fixed point problem $f=Tf$ where the
non-linear map $T$ preserves the space of positive, monotonously
decreasing functions. With some effort it seems possible to prove that
$T$ preserves the space of completely monotonic functions. It is of
course not obvious that complete monotonicity of $a\mapsto G(a,0)$ is
transferred to $a\mapsto G(a,a)$ given by (\ref{Gab}).

In between the Stieltjes functions and the completely monotonous
functions lies the class of generalised Stieltjes functions of order
$\kappa>0$ which admit a representation 
\begin{align}
f(x)= c+\int_0^\infty \frac{d \rho_\kappa(M^2)}{(x +M^2)^\kappa}\;, 
\end{align}
where $c \geq 0$ and $\rho_\kappa(M^2)$ is a positive measure.  For a
review on generalised Stieltjes functions we refer to a recent article
of Sokal \cite{Sokal:2010??}  which identifies the precise conditions
under which a real function $f$ is a generalised Stieltjes function.
The identity \cite[eq.~(7)]{Sokal:2010??}
\begin{align}
\frac{1}{(x+t)^\kappa}=\frac{\Gamma(\kappa')}{\Gamma(\kappa)
\Gamma(\kappa'-\kappa)}
\int_0^\infty du \;u^{\kappa'-\kappa-1} \frac{1}{(x+t+u)^{\kappa'}}
\label{Sokal}
\end{align}
implies that a Stieltjes function of order $\kappa$ is also a
Stieltjes function of order $\kappa'>\kappa$. Moreover, any generalised
Stieltjes function is completely monotonic. 

If $a \mapsto G(a,a)$ happens to be a generalised Stieltjes function
of order $2 \leq k\in \mathbb{N}$ (integer-order suffices by
(\ref{Sokal})), then the same steps as in the proof of
Proposition~\ref{Prop:Stieltjes} yield an analytic continuation of the
Schwinger function $S_2$ to Minkowski space. The big difference is
that the corresponding Wightman function $\hat{W}^1_2(q)$ involves the
derivative $\delta^{(k-1)}(\frac{(q^0)^2-\vec{q}\cdot
  \vec{q}-M^2}{\mu^2})$ of the Dirac distribution which is not
positive for $k\geq 2$. This means that \emph{Osterwalder-Schrader
  reflection positivity cannot be expected for $a\mapsto G(a,a)$ being
  a generalised Stieltjes function of order $\kappa >1$}.

Numerical investigations \cite{GW-numerical} and also the perturbative
solution of the 2-point function $G(a,a)$ tend to suggest the
asymptotic behaviour
\begin{align}
G(a,a) \stackrel{a\to \infty}{\propto} 
\frac{1}{(1+2(1+\mathcal{Y})a)^{1+\lambda}}\;.
\label{Gaa-asymp}
\end{align}
This would imply that for the physical coupling constant $\lambda>0$
the matrix 2-point function $a\mapsto G(a,a)$ is not Stieltjes and as
such does not permit an analytical continuation to a positive Wightman
quantum field theory. A related consequence is the negative anomalous
dimension for $\lambda>0$: If (\ref{Gaa-asymp}) holds exactly, i.e.\
$G(\frac{\|p\|^2}{2\mu^2(1+\mathcal{Y})}, \frac{\|p\|^2}{2\mu^2
  (1+\mathcal{Y})})= \mbox{$(\frac{\|p\|^2}{\mu^2}+1)^{-(1+\lambda)}$}$, then
(\ref{S-2}) becomes $S_c(\mu x,\mu y) =\frac{2^{-\lambda}}{4\pi^2
  \Gamma(1+\lambda)}
\frac{K_{1-\lambda}(\mu\|x-y\|)}{(\mu\|x-y\|)^{1-\lambda}}$. To obtain
this result one expresses
\mbox{$(\frac{\|p\|^2}{\mu^2}+1)^{-(1+\lambda)}$} by the
$\Gamma$-function integral, evaluates the resulting Gau\ss{}ian
integral over $p\in \mathbb{R}^4$ and uses the integral representation
\cite[\S 8.432.6]{Gradsteyn:1994??} of the modified Bessel function
$K_\nu(z)$. From $K_\nu(z)\stackrel{z\to 0}{\propto}
\frac{\Gamma(\nu)}{2} (\frac{2}{z})^\nu$ \cite[\S
9.6.9]{Abramowitz:1972??} it follows $S_c(\mu x,\mu y)\stackrel{x -y
  \to 0}{\propto} \frac{2^{-2\lambda} \Gamma(1-\lambda)}{4\pi
  \Gamma(1+\lambda)} \frac{1}{(\mu\|x-y\|)^{2-2\lambda}}$, which means
that the anomalous dimension would be $\eta=-2\lambda$.

  Conversely, (\ref{Ga0}) and (\ref{Gab}) might\footnote{Existence of
    a solution of (\ref{Ga0}) is, so far, only established for
    $\lambda>0$.}  \emph{define} an analytical continuation of the
  model to \mbox{$\lambda<0$}. This wrong-sign noncommutative
  $\lambda\phi^4_4$-model could then have an analytical continuation
  to a Wightman quantum field theory. In a certain sense this
  parallels the construction of the commutative planar wrong-sign
  $\lambda\phi^4_4$-model by t'Hooft \cite{'tHooft:1982cx} and
  Rivasseau \cite{Rivasseau:1983jj}.

The negative anomalous dimension resulting from the faster decay of 
$G(a,a)$ in $a$ for $\lambda>0$ in comparison with the free
theory $\lambda=0$ where $G(a,a)=\frac{1}{1+2a}$ exactly is the result
of the renormalisation. The two-dimensional model which does not
require a wavefunction renormalisation has, at least perturbatively,
the opposite behaviour
$G(a,a)^{(D=2)}=\frac{1}{1+2a}+\frac{\lambda}{\mu^2} 
\frac{2\log(1+a)}{(1+2a)^2}+\mathcal{O}(\lambda^2)$ than the
4-dimensional case 
$G(a,a)^{(D=4)}=\frac{1}{1+2a}-\lambda
\frac{(2+2a)\log(1+a)}{(1+2a)^2}+\mathcal{O}(\lambda^2)$. The
perturbative result also suggests that in $D=2$ 
the difference between free and interacting theory is
subleading to a power law, $G(a,a)^{(D=2)} 
\stackrel{a\to \infty}{\propto} \frac{1}{1+2a}$
independent of the coupling constant. This favours the conjecture that
the two-dimensional model can define a Wightman theory for any sign of
the coupling constant.

\section{Interpretation}

In this paper we have translated the matrix model correlation functions
of Moyal-deformed $\phi^4_4$-theory solved in \cite{Grosse:2012uv} to
position space. This involves a different infinite volume limit as
pointed out below:

\begin{enumerate} 
\item \emph{Matrix model limit.\quad} This limit arises directly in
  matrix formulation \cite{Grosse:2012uv} from the free energy density
  $\mathcal{W}=\lim_{V\to \infty}\frac{1}{V}\log\mathcal{Z}$ with its
  usual volume dependence.  As seen from
  \cite[Prop.~3.5]{Grosse:2012uv}, this limit eliminates all
  non-trivial topology of the matrix model, i.e.\ both the non-planar
  sector with genus $g\geq 1$ and the sector with $B\geq 2$ boundary
  components. The restriction to trivial topology agrees with other
  large-$\mathcal{N}$ limits of matrix models. There remain the planar
  regular $N$-point functions $G(b_0,\cdots ,b_{N-1})$ with $b_i \in
  \mathbb{R}_+$ for which there is an exact recursion formula
  \cite[eq.\ (4.50)]{Grosse:2012uv} that provides $G(b_0,\cdots,
  b_{N-1})$ as weighted difference quotients of products of two-point
  functions $G(a,b)$. The combinatorics involves non-crossing
  partitions counted by the Catalan numbers. The two-point function is
  given as a function (\ref{Gab}) of its boundary $G(a,0)$, which
  itself is the solution of a non-linear integral equation
  (\ref{Ga0}).

\item \emph{Statistical physics limit.\quad} This is the limit which
  gives the Schwinger functions of Definition~\ref{Def:Schwinger}.
  There are two non-na\"{\i}ve volume scalings involved, a procedure
  that is common in statistical physics. We first define the free
  energy density as $\mathcal{F}=\frac{1}{(V\mu^4)^2}\log\mathcal{Z}$.
  The additional volume factor is due to the fact that the spectral
  geometry \cite{Grosse:2007jy, Gayral:2011vu} behind the
  noncommutative quantum field theory under consideration has a finite
  volume $V^2$, not $V$. According to the second and third remark
  after Definition~\ref{Def:Schwinger}, there is also a wavefunction
  renormalisation $\sqrt{Z}$ to $\frac{\sqrt{Z}}{V\mu^4}$ involved.

According to \cite[Prop.~3.5]{Grosse:2012uv}, $\mathcal{F}$ has an
expansion into planar topological sectors with $B$ boundary
components and prefactor $\frac{1}{(V\mu^4)^B}$.  The next step is
to notice that individual matrix element correlation functions give
according to (\ref{phi-x-fmn}) and (\ref{fmn}) the amplitude of a
Gau\ss{}ian wave packet in position space. The assembly of plane waves
from Gau\ss{} packets involves sums over the matrix indices. As proved
by Corollary~\ref{Cor:Laguerre}, this index summation produces a
factor $V\mu^4$ per boundary component with even length, whereas no
such factor arises for a boundary component of odd length. This means
that all sectors with $B\geq 1$ and an even number of sources/fields
per boundary component contribute to $\mathcal{F}$ in \emph{position
  space}. This makes the statistical physics limit the topologically
richest one.
\end{enumerate}

The resulting Schwinger functions $S(\mu x_1,\dots,\mu x_N)$ given in
(\ref{Schwinger-final}) define a Euclidean quantum field theory on
$\mathbb{R}^4$. These Schwinger functions have the full Euclidean
invariance, they are symmetric and (as discussed in
section~\ref{sec:continuation}) they might possess an analytic
continuation to Wightman functions of a four-dimensional relativistic
quantum field theory, possibly only for $\lambda \leq 0$.  The
resulting field theory limit is close to a free theory, but there are
differences that we describe below.

The richest sector of the model is the case $B=\frac{N}{2}$ made of
the $(2{+}2{+}\dots{+}2)$-point functions. This sector describes the
propagation and interaction of $B=\frac{N}{2}$ particles with momenta
$p_1,\dots,p_{\frac{N}{2}}$.  These particles interact, but in a way
that the \emph{momentum is unchanged}, precisely as with free fields.
If two or more of these $\frac{N}{2}$ particles have coinciding
momenta, then another interaction channel is opened which is described
by the sectors with $B< \frac{N}{2}$. It would be interesting to 
extend this model to scalar fields of several components. In this case
momentum could be exchanged between the components.  

The key difference to a free theory is the (maximal) violation of the
cluster property. Any connected ($N{\geq} 4$)-point Schwinger function
(\ref{Schwinger-final}) contains contributions which do not decay to
zero if a subset of positions $x_i$ is shifted infinitely away. Let us
consider the case $N=4$ in (\ref{Schwinger-final}). For $0\neq x\in
\mathbb{R}^4$ one has 
\begin{align}
&\lim_{\tau\to \infty} 
S_4(\mu x_1,\mu x_2 ,\mu(x_3+\tau x),\mu(x_4+\tau x))
\nonumber
\\*
&= 
\int_{\mathbb{R}^4\times \mathbb{R}^4} \frac{d^4 p \,d^4q}{2\pi^6 \mu^8}
e^{\mathrm{i} \langle p, x_1-x_2\rangle + 
\mathrm{i} \langle q, x_3-x_4\rangle }
G\Big(\tfrac{\|p\|^2}{2\mu^2(1+\mathcal{Y}) },
\tfrac{\|p\|^2}{2\mu^2(1+\mathcal{Y})} \Big|
\tfrac{\|q\|^2}{2\mu^2(1+\mathcal{Y})} ,
\tfrac{\|q\|^2}{2\mu^2(1+\mathcal{Y})} \Big)
\nonumber
\\*
&+ 
\int_{\mathbb{R}^4} \frac{d^4 p}{(2\pi \mu)^4}
\frac{e^{\mathrm{i} \langle p, x_1-x_2
+ x_3-x_4\rangle }
+  e^{\mathrm{i} \langle p, x_1-x_2
+ x_4-x_3\rangle }}{2} 
G\Big(\tfrac{\|p\|^2}{2\mu^2(1+\mathcal{Y}) },
\tfrac{\|p\|^2}{2\mu^2(1+\mathcal{Y})} ,
\tfrac{\|p\|^2}{2\mu^2(1+\mathcal{Y})} ,
\tfrac{\|p\|^2}{2\mu^2(1+\mathcal{Y})} \Big)\;,
\end{align}
because all other permutations vanish almost everywhere 
by the Riemann-Lebesgue lemma.

Assuming validity of the other Osterwalder-Schrader axioms
\cite{Osterwalder:1973dx, Osterwalder:1974tc}, the corresponding
Wightman quantum field theory would also lack the clustering property.
Wightman's reconstruction theorem \cite{Streater:1964??} then implies
that the vacuum would be a mixed state. Its decomposition into pure
states corresponds to a decomposition into different topological
sectors. It would be very interesting to study this non-trivial
topology of the $\phi^4_4$-model on Moyal space in the limit
$\theta\to \infty$. Unfortunately, the lack of more detailed knowledge
about the diagonal matrix 2-point function $G(a,a)$ moves this
investigation into the future.

At first glance it seems surprising that the limit $\theta\to\infty$
of infinite noncommutativity is so close to a traditional quantum
field theory expected for $\theta\to 0$. An intuitive explanation is
the following. The Feynman rule in momentum space for the quartic
interaction vertex with momenta $p_1,\dots,p_4$ reads
$\mathcal{V}(p_1,\dots,p_4)=\lambda e^{\mathrm{i}\sum_{i<j} p_i^\mu
  \theta_{\mu\nu} p_j^\nu}\delta(p_1+\dots+p_4)$. If
$f(p_1,\dots,p_4)$ is any $L^1$-function of the momenta, then
$\lim_{\theta\to \infty} f(p_1,\dots,p_4)\mathcal{V}(p_1,\dots,p_4)=0$
almost everywhere by the Riemann-Lebesgue lemma.  This shows that up
to measure zero, the $\theta\to \infty$ limit of a quantum field
theory on Moyal space is a free theory.  To the exceptional points
where the limit is non-zero belong linearly dependent momenta $p_i$
where the phase vanishes.  Now comes the crucial point: \emph{These
  subspaces of total measure zero where the theory is not free are
  possibly protected for topological reasons, and this is the case for
  $B>1$ boundary components}. The subspace where the momenta of fields
attached to each boundary add up to zero has full Lebesgue measure. In
connection with the correct volume-dependent field renormalisation
this establishes qualitatively our result that \emph{the $\theta\to
  \infty$ limit of noncommutative $\phi^4_4$-theory differs from a
  free theory by the presence of non-trivial topological sectors}.

\section*{Acknowledgements}

This work was initiated at the Erwin Schr\"odinger Institute in Vienna
which financed a 6-week visit of one of us (RW). The first steps
were jointly achieved with Gandalf Lechner whose visit to Vienna was
financed by an FWF project of Jakob Yngvason. We would like to
cordially thank Gandalf Lechner for these fruitful discussions. We
would also like to thank Vincent Rivasseau for pointing out to us the
relation between asymptotics of the 2-point function and reflection
positivity.

\begin{appendix}
\section{Sum over products of Laguerre polynomials}

\begin{Lemma}
For $t_i\in \mathbb{R}_+$, $z_i\in \mathbb{C}$ with $|z_i|<1$ and 
cyclic identification $J+i\equiv i$ of indices one has
\begin{align}
\sum_{m_1,\dots,m_J=0}^\infty 
\prod_{i=1}^J z_i^{m_i}\, L_{m_i}^{m_{i+1}-m_i}(t_i)  
&=\frac{1}{1-(z_1\cdots z_J)}
\exp\Bigg(-\frac{
\mbox{\small$\displaystyle\sum_{i,j=1}^Jt_i (z_{j+i}\cdots z_{J+i})$}
}{1-(z_1\cdots z_J)}
\Bigg)\;.
\label{sum-Laguerre-1}
\end{align}
\end{Lemma}
\emph{Proof.}  We use the generating function \cite[\S
8.975.3]{Gradsteyn:1994??} of Laguerre polynomials
\begin{align}
\sum_{n=0}^\infty L_n^{\alpha -n}(t)z^n=e^{-zt} (1+z)^\alpha
\label{Laguerre-gen}
\end{align}
to split up the index chain: 
\begin{align}
&\sum_{m_1,\dots,m_J=0}^\infty 
\prod_{i=1}^J z^{m_i} L^{m_{i+1}-m_i}_{m_i}(t_i)
\nonumber
\\
&= \sum_{m_1,\dots,m_J=0}^\infty 
\Big(\prod_{i=1}^{J-1} z_i^{m_i} 
L^{m_{i+1}-m_i}_{m_i}(t_i)\Big)\cdot 
\frac{1}{m_J!} \frac{d^{m_J}}{du^{m_J}} \sum_{n=0}^\infty 
u^n z_J^{n} L^{m_{1}-n}_{n}(t_J)\Big|_{u=0}
\nonumber
\\
&= \sum_{m_1,\dots,m_J=0}^\infty 
\Big(\prod_{i=1}^{J-1} z_i^{m_i} 
L^{m_{i+1}-m_i}_{m_i}(t_i)\Big)\cdot 
\frac{1}{m_J!} \frac{d^{m_J}}{du^{m_J}} e^{-u z_J t_J}
(1+uz_J)^{m_1}\Big|_{u=0}\;.
\end{align}
Now we successively sum over $m_1,m_2,\dots,m_{J-1}$ where each sum is
of the form (\ref{Laguerre-gen}). There remains a final sum over
$m_J\equiv m$:
\begin{align*}
\sum_{m_1,\dots,m_J=0}^\infty 
\prod_{i=1}^J  z_i^{m_i} L^{m_{i+1}-m_i}_{m_i}(t_i)
= \sum_{m=0}^\infty \frac{1}{m!} \frac{d^m}{du^m}
\Big(\Big(A+u\prod_{i=1}^J z_J\Big)^me^{-uB-C}\Big)\Big|_{u=0}
\;, \hspace*{-0.7\textwidth}
\nonumber
\\*
A&:=1+\sum_{i=1}^{J-1} z_i\cdots z_{J-1}
&
B&:=\sum_{i=0}^{J-1} 
\big(z_0 \cdots z_i\big) t_i\;,
&
C&:=\sum_{i=1}^{J-1} \sum_{j=1}^i (z_j\cdots z_i)t_i \;.
\end{align*}
This leads to
\begin{align}
\sum_{m_1,\dots,m_J=0}^\infty 
\prod_{i=1}^J  z_i^{m_i} L^{m_{i+1}-m_i}_{m_i}(t_i)
&=e^{-C} \sum_{m=0}^\infty \sum_{k=0}^m\frac{1}{m!} \binom{m}{k} 
\binom{m}{k} (m-k)! (-BA)^k 
\Big(\prod_{i=1}^J z_i\Big)^{m-k}
\nonumber
\\
&= e^{-C} \sum_{k=0}^\infty \sum_{m=0}^\infty\frac{(m+k)!}{k!k!m!} 
(-BA)^k
\Big(\prod_{i=1}^J z_i\Big)^{m}
\nonumber
\\
&= \frac{1}{1-\prod_{i=1}^J z_i} 
\exp\Big(-C-\frac{BA}{1-\prod_{i=1}^J z_i} \Big)\;.
\end{align}
It is straightforward to check
$\displaystyle 
C+\frac{BA}{1-\prod_{i=1}^J z_i}= \frac{
\mbox{\small$\displaystyle\sum_{i,j=1}^J t_i (z_{j+i}\cdots z_{J+i})$}
}{1-(z_1\cdots z_J)}$, 
which yields the assertion (\ref{sum-Laguerre-1}). \hfill $\square$%

\bigskip

\begin{Corollary}
\label{Cor:Laguerre}
Let $\langle x,y\rangle$, $\|x\|$ and
  $x{\times}y=\det(x,y)$ be scalar product, norm and (third component
  of) vector product of $x,y\in \mathbb{R}^2$. Then for $x_i \in
  \mathbb{R}^2$ and $z_i\in \mathbb{C}$ with $|z_i|<1$, the
$f_{mn}(x_i)$ defined in (\ref{fmn}) satisfy
(with cyclic identification $J+i\equiv i$ of indices where necessary)
\begin{align}
&\sum_{m_1,\dots,m_J=0}^\infty \prod_{i=1}^J 
f_{m_im_{i+1}}(x_i) z_i^{m_i}
\nonumber
\\
&=\frac{2^J}{1-\prod_{i=1}^J (-z_i)} 
\exp\bigg(-\frac{\sum_{i=1}^J \|x_i\|^2}{\theta}
\frac{1+\prod_{i=1}^J (-z_i)}{1-\prod_{i=1}^J (-z_i)}\bigg)
\nonumber
\\
& \times \exp\bigg(\!\!\!-\! \frac{2}{\theta}
\!\!\! \sum_{1\leq k < l \leq J} \!\!\! 
\Big(\!
\big(\langle x_k, x_l\rangle
{-}\mathrm{i} x_k{\times} x_l\big)\frac{\prod_{j=k+1}^l ({-}z_j)}{
1{-}\prod_{i=1}^J ({-}z_i)} 
+
\big(\langle x_k, x_l\rangle
{+}\mathrm{i} x_k{\times} x_l\big)
\frac{\prod_{j=l+1}^{J+k} ({-}z_j)}{1{-}\prod_{i=1}^J ({-}z_i)} 
\! \Big)\!\bigg).
\label{sum-Laguerre}
\end{align}
\end{Corollary}
\emph{Proof.}  
{}From (\ref{fmn}) we get for
$0\neq x_i \in\mathbb{R}^2\equiv \mathbb{C}$ and $|z_i|$
sufficiently small 
\begin{align}
\prod_{i=1}^J 
f_{m_im_{i+1}}(x_i) z^{m_i}
&=2^J e^{-\frac{1}{\theta} \sum_{i=1}^J x_i \overline{x_i}} 
\prod_{i=1}^J \tilde{z}_i^{m_i} 
L^{m_{i+1}-m_i}_{m_i}(t_i)\Big|_{t_i=\frac{2}{\theta}x_i
  \overline{x_i}, ~
\tilde{z}_i=-z_i \frac{x_{i-1}}{x_i}} \;.
\end{align}
This yields
\[
\sum_{i,j=1}^J t_i (\tilde{z}_{j+i}\cdots \tilde{z}_{J+i})
= \frac{2}{\theta} 
\sum_{i,j=1}^J x_{i+j-1}\overline{x}_i 
(-z_{j+i})\cdots (-z_{J+i})
\]
for the sum (\ref{sum-Laguerre-1}) of Laguerre polynomials.  Splitting
the sum into the cases $j=1$, $j=2,\dots J-i$ and $j=J-i+1 \dots J$ we
confirm (\ref{sum-Laguerre}) with $x_i\overline{x_j}=\langle
x_i,x_j\rangle - \mathrm{i} x_i\times x_j$. \hfill $\square$%

\end{appendix}

\end{document}